\documentclass[11pt,a4paper]{article}
\usepackage[utf8]{inputenc} %Para las tildes en formato UTF-8. !Instalé latex-ucs
\usepackage[english]{babel}
\usepackage{graphicx,amssymb,amsmath,wasysym,subfigure}
\usepackage{epstopdf}
\usepackage[bookmarks=false]{hyperref}
 
\def\be{\begin{equation}}
\def\beqn{\begin{eqnarray}}
\def\ee{\end{equation}}
\def\eeqn{\end{eqnarray}}
\def\mhalf{m_{1/2}}
\def\mhu{m_{H_U}^{2} (M_{GUT})}
\def\mhd{m_{H_D}^{2} (M_{GUT})}
\def\tanb{\tan\beta}
\def\neut1{\chi^0_1}
\def\issue(#1,#2,#3){{\bf #1}, #2 (#3)}
\def\PREP(#1,#2,#3){Phys.\ Rep. \issue(#1,#2,#3)}

\begin{document}
\baselineskip=15.5pt

\thispagestyle{empty}

\begin{flushright}
LPT-10-52
\end{flushright}
%\centerline{\bf \today}

\vspace{.5cm}

\begin{center}

{\Large\sc{\bf 
%SUSY funnel region in the MSSM Light Higgs scenario : Possibilities in the upcoming Dark Matter detection experiments. 
Exploring SUSY light Higgs boson scenarios via dark matter experiments.}}
%\vskip 0.3 cm
%{\Large\sc{\bf 
%SUSY funnel region in the MSSM Light Higgs scenario : Possibilities in the upcoming Dark Matter detection experiments. 
%}}

\vspace*{9mm}
\setcounter{footnote}{0}
\setcounter{page}{0}
\renewcommand{\thefootnote}{\arabic{footnote}}

\mbox{ {\sc Debottam Das$^{1}$, Andreas Goudelis$^{1}$, Yann Mambrini$^{1}$}}

\vspace*{0.9cm}

{\it $^1$Laboratoire de Physique Théorique d'Orsay, UMR8627--CNRS,\\
Université Paris--Sud, Bât. 210, F--91405 Orsay Cedex, France}\\

\vskip5mm
E-mails: {\tt debottam.das@th.u-psud.fr, andreas.goudelis@th.u-psud.fr, yann.mambrini@th.u-psud.fr}
\vskip5mm

\end{center}

\vspace{1cm}

\begin{abstract}
We examine the dark matter phenomenology in supersymmetric 
light higgs boson scenarios, adapting nonuniversal Higgs masses at the gauge coupling unification scale. The correct relic density
is obtained mostly through the annihilation into a pseudoscalar $A$, 
which gives high values for the self-annihilation cross-section 
at present times.
Our analysis shows that most part of the $A$ pole region can produce
detectable gamma-rays and antiproton signals, and still be compatible with 
with recent direct detection data from XENON100 and CDMS-II.
\end{abstract}

\newpage

\tableofcontents

\newpage

%%%%%%%%%%%%%%%%%%%%%%%%%%%%%%%%%%%%%%%%%%%%%%%%%%%%%%%%%%%%%%%%%%%%%%%%%%%%%%%%%%%%%%%%%%%%%%%%%%%%%%%
%%%%%%%%%%%%%%%%%%%%%%%%%%%%%%%%%%%%%%%%%%%%%%%%%%%%%%%%%%%%%%%%%%%%%%%%%%%%%%%%%%%%%%%%%%%%%%%%%%%%%%%
%%%%%%%%%%%%%%%%%%%%%%%%%%%%%%%%%%%%%%%%%%%%%%%%%%%%%%%%%%%%%%%%%%%%%%%%%%%%%%%%%%%%%%%%%%%%%%%%%%%%%%%
\section{Introduction}
\label{sec:intro}

Low energy supersymmetry (SUSY) \cite{SUSY}, 
has emerged as the most promising 
candidate for physics beyond the Standard Model (SM). 
The minimal supersymmetric standard model (MSSM) has been attractive for its  
explanations towards several phenomena such as~(i) the Higgs mass hierarchy problem in the SM, 
(ii)~electroweak (EW) symmetry breaking via 
radiative corrections and (iii)~the unification of the SM gauge couplings at the grand 
unification scale ($M_{GUT} \simeq 2\times 10^{16}$~GeV). Assuming R-parity
conservation, a specially attractive
feature of the MSSM is the presence of the stable, weakly interacting lightest
supersymmetric particle (LSP) in the form of lightest neutralino 
(${\chi}_1^0$) \cite{goldbergDM} 
which turns out to be a good candidate for the observed cold dark matter (CDM) in the universe.

However, the MSSM has a large number of soft breaking 
parameters, which make it less predictive in its most general forms.  
Most of these 
parameters can be elliminated by considering a definite SUSY breaking
framework at the high scale. The simplest model in this regard, which
assumes gravity as the mediator for SUSY breaking - the minimal supergravity 
(mSUGRA) \cite{msugra} model - has only five free parameters.
These are the common gaugino and scalar 
mass parameters $m_{1/2}$ and $m_0$, the common tri-linear coupling 
parameter $A_0$, all given at the gauge coupling unification scale,
 the ratio of Higgs vacuum expectation 
values at the electroweak scale, namely $\tan\beta$, and the sign of 
the Higgsino mixing 
parameter $\mu$. The lightest 
neutralino can be expressed in terms of its gauge eigenstates, the 
Bino, the Wino and the two Higgsinos as
\be
{\chi^0_1} \equiv N_{11} \tilde B +N_{12} \tilde W+N_{13} \tilde H_D+
N_{14} \tilde H_U.
\label{lspcompositioneqn}
\ee  
The model is primarily associated with 
a Bino-dominated LSP that leads to over-abundance of dark matter. 
Reduction of the relic density to satisfy the 
WMAP data can be achieved if i) there is 
coannihilation of the LSP with another sfermion (usually stau ${\tilde \tau}_1$ 
or rarely ${\tilde t}_1$ ) that has a mass 
close to the mass of the LSP \cite{coannistau}, ii) there is appropriate mixture of 
Bino and Higgsinos in the composition of the LSP 
so that there may be coannihalating charginos 
in the LSP-${\tilde \chi}_1^\pm$ annihilation \cite{Edsjo:1997bg,mizuta}
(the $\chi_1^0 \chi_1^0 Z$ coupling is also enhanced),  
the so called 
focus point (FP) \cite{focus}/hyperbolic branch (HB) \cite{hyper} region 
(this kind of coannihilation
becomes significant when the LSP is dominantly Higgsino like),     
 iii) the LSP is 
sufficiently small in mass and sfermions are light so that light sfermion 
exchange may enhance the LSP-LSP annihilation rates
{\footnote{This region known as {\it bulk annihilation} region, 
though highly constrained by the LEP2 limits on the Higgs boson mass, 
could be revived with non-vanishing $A_0$ parameter \cite{my3}.}}, 
or 
iv) there is a possibility of having s-channel Higgs exchanges 
occuring via the CP-odd Higgs boson 
$A$ or via CP-even heavy (light) Higgs bosons $H$ ($h$) leading to the 
``funnel region'' of the dark matter satisfying zone in 
the $m_{1/2}-m_0$ plane \cite{funnel,djouadiLightHiggs}. 

This simple model is highly predictive because of its small number of
unknown parameters. 
However, a very severe restriction
 appears on the $(m_{1/2}-m_0)$ plane of mSUGRA from
the LEP2 bound on lightest Higgs boson mass ($m_h$) \cite{hlim}, which reads 
\be
                   m_h> \rm 114.4 \ GeV.
\label{higgs}
\ee

\noindent           
In order to satisfy the limit $m_h >$114 GeV within the framework of mSUGRA, one 
requires reasonably large values for the
top squark mass ($\gg M_Z$) which implies some fine-tuning of SUSY parameters to
obtain the correct value of $M_Z$.
However, the limit of Eq.\eqref{higgs} is related to the principle Higgs search channel
in the LEP2 that involves production of the Higgs boson via the 
$e^+e^- \rightarrow Zh$ process, followed by decays of the Higgs boson.  
The production process principally depends on the coupling $g_{ZZh}$. 
Now, in the MSSM, the coupling $g_{ZZh}$ is not identical
with the respective coupling in the SM, it can rather be written as 
${\sin}(\beta-\alpha)g^{\it SM}_{ZZh}$(For reviews see 
 \cite{HaberNP,djouadihiggs}). 
Here, $g^{\it SM}_{ZZh}$ represents
the SM Higgs coupling to the $Z$ boson. The prefactor 
${\sin}(\beta-\alpha)$ appears due to the mixing of the CP-even Higgs bosons,
where $\alpha$ is the mixing
angle. In supersymmetric theories 
the limit of Eq.\eqref{higgs}
refers to the scenarios where ${\sin}(\beta-\alpha) \sim 1$. This corresponds to
the so-called
{\it decoupling} zone for the Higgs boson where one finds $m_A^2 >> M_Z^2$. 
In fact this 
is true for most of the parameter points in mSUGRA. However, in the general MSSM
the above prefactor may become much smaller. This opens up a 
supersymmetric parameter space associated
with a Higgs mass smaller than the limit of Eq.\eqref{higgs} (For details see 
 \cite{djouadihiggs}).
This is the
{\it non decoupling} zone for the Higgs boson, where all Higgs masses may have
comparable values $m_h \simeq m_H \simeq m_A \sim M_Z$
 \cite{Drees:2005jg,kane,Belyaev-LHS,kim,hisano,Cao:2010ph,boosetal,yannintense}.
An important virtue is that with this new window, a light Higgs mass with 
$m_h \sim (95-100)$
GeV could explain the slight excess of events of $2.3\sigma$ statistical 
significance  \cite{Drees:2005jg} 
as reported by the LEP2 experiments  \cite{hlim}. Within this 
{\it non decoupling} zone
 it is possible to have a Higgs boson lighter than the 
$Z$ boson  \cite{kane,Belyaev-LHS}. In this regime, the Higgs mass does not require large radiative
corrections, thus
could evade the little hierarchy problem  \cite{kim}.

This particular region, which has recently been named as 
'{\it Light Higgs Boson}' scenario
or LHS  \cite{Belyaev-LHS,hisano}, 
is also interesting as it provides definite 
predictions.  Since all Higgs states are light enough, they can be produced
at the Large Hadron Collider (LHC)  \cite{Belyaev-LHS}.
Interestingly, this region can also accommodate the WMAP  
data for the dark matter relic density   \cite{hisano}. 
However, realization of the {\it light Higgs boson} scenario
is not possible in the most constrained model like mSUGRA. 
Scenarios like supergravity - inspired nonuniversal Higgs mass models (NUHM), 
where the Higgs mass square parameters  
may have different values at the scale $M_{GUT}$ 
($m_{H_U}^{2} (M_{GUT}), m_{H_D}^{2} (M_{GUT})$)
from the other scalars  \cite{NUHM1,NUHM2}, 
may be useful in this regard. Similarly,
this particular SUSY zone can  be realized in a 
particular nonuniversal scalar mass model where third generation squarks and
the two Higgs doublets may have vanishing values at the GUT scale  \cite{mynusm}. 

In this work, adapting a NUHM framework, we try to explore 
this particular parameter space 
in the light of recent dark matter detection experiments. 
Moreover, our focus will be to 
compute the indirect detection rate, particularly where the mass 
of the lightest neutralino and 
Higgs bosons are reasonably small. 
We will try to find out the parameter space where the WMAP data is satisfied 
via the resonant annihilation into a CP-odd Higgs boson. This is
interesting as then neutralino pair annihilations in our galaxy could
produce detectable signals both in the antiproton and the gamma-ray channels.
In fact, the part of the MSSM parameter space with
light neutralino masses which is otherwise forbidden in models like mSUGRA, 
can completely be explored in the near future. 
The other WMAP - satisfying zones, as mentioned, 
% are not quite friendly in achieving the 
cannot produce such
large anti-proton/gamma ray signals (except the FP/HB zone where 
the Bino dominated LSP with sufficient Higgsino component could lead to
large signals). 
Defining 
a measure for these observables, we examine the detection prospects for the 
parameter
space in the ongoing as well as upcoming experiments.

Now, it is not surprising that the neutralino - nucleus spin-independent
scattering cross-section
$\sigma^{SI}$, which is dominated by $t$-channel Higgs exchange ($h,H$) 
could be very large in the present context. The cross-section 
reaches its maximum value in the {\it non decoupling} zone because of smaller 
mass values of the CP-even Higgs bosons. 
In fact this particular region may fall within the range of the exclusion 
limits as given
by CDMS-II \cite{cdms-2} or XENON100 \cite{xenon100}. 
However there exist considerable
uncertainties in the computation of the above cross-section as well as the
relevant experimental limits. 
The principle sources are related to
(i) nuclear form factors, 
(ii) the WIMP - nucleon scattering cross-section (mainly concerning
the strange quark content of the nucleon), 
(iii) the local dark matter density
and (iv) the escape velocity of the WIMP. 
Considering all these uncertainties 
and particularly exploiting the uncertainty related to the strange quark 
content of the nucleon we will see that the predicted
neutralino-nucleus scattering cross-section $\sigma^{SI}$ may change 
significantly. 
These uncertainties render the question of exclusion of interesting 
regions of the parameter space
related to dark matter phenomenology subtle to answer.

  The paper is organized as follows.  In Sec.\ref{sec:model} we will primarily 
discuss the model and the low energy constraints which are relevant here. In Sec.\ref{dir_detection}
we introduce the spin-independent neutralino-nucleus scattering cross-section for the LSP. Similarly, 
Sec.\ref{ind_detection} is devoted to the indirect detection of the LSP, primarily to the (i) gamma-ray
and the (ii) anti-proton signal. We describe our results in 
Sec.\ref{resultsall}. In Sec.\ref{sec:DirectDetection}
we provide estimates regarding the effect of the uncertainties in the 
computation of $\sigma^{SI}$. Finally we conclude in Sec.\ref{sec:con}.

%%%%%%%%%%%%%%%%%%%%%%%%%%%%%%%%%%%%%%%%%%%%%%%%%%%%%%%%%%%%%%%%%%%%%%%%%%%%%%%%%%%%%%%%%%%%%%%%%%%%%%%
%%%%%%%%%%%%%%%%%%%%%%%%%%%%%%%%%%%%%%%%%%%%%%%%%%%%%%%%%%%%%%%%%%%%%%%%%%%%%%%%%%%%%%%%%%%%%%%%%%%%%%%
%%%%%%%%%%%%%%%%%%%%%%%%%%%%%%%%%%%%%%%%%%%%%%%%%%%%%%%%%%%%%%%%%%%%%%%%%%%%%%%%%%%%%%%%%%%%%%%%%%%%%%%
\section{The NUHM and low energy constraints}
\label{sec:model}
In this section, we describe the model parameters and the relevant low energy
constraints. As mentioned in the introduction, we place ourselves in a 
supergravity framework 
with nonuniversal Higgs masses at the GUT scale. Thus, the model can be 
specified by the following parameters:
\be
\mhalf,A_0,m_0,\mhu,\mhd,sign(\mu),\tan\beta.
\ee
Here, $\mhalf$ and $A_0$ represent universal values for the gaugino masses
and the trilinear coupling parameters, while
$m_0$ denotes universal masses for all
scalars except the Higgs bosons at the GUT scale. 
In general, Higgs scalars can belong to
different multiplets of a grand unified gauge group, thus may assume different
values than the other scalars in the theory
(see \cite{String} for motivated constructions leading to specific nonuniversal
Higgs mass terms).
 In this
spirit, we scan over the Higgs mass square parameters 
$\mhu$ and $\mhd$ at the scale $M_{GUT}$  
to produce a MSSM 
parameter space consistent with all low energy constraints
and where Higgs bosons can be reasonably lighter. 
Using the Renormalization
Group Equations (RGE) for the soft SUSY breaking parameters along with the 
conditions for radiative EWSB one can determine $\mu$ and $m_A$ through
\beqn
\mu^2 = 
-\frac{1}{2} M^2_Z +\frac {m_{H_D}^2-m_{H_U}^2 \tan^2\beta} {\tan^2\beta -1}
+ \frac {\Sigma_1 -\Sigma_2 \tan^2\beta} {\tan^2\beta -1}
\label{rewsbmusqeqn}
\eeqn
and
\beqn
\sin2\beta =  2B\mu/(m_{H_D}^2+m_{H_U}^2+2\mu^2+\Sigma_1+\Sigma_2)
\label{rewsbBeqn}
\eeqn
where the $\Sigma_i$'s represent the one-loop 
corrections~ \cite{effpot1,effpot2}, which become small at the 
scale where the Higgs potential $V_{Higgs}$ is minimized.
We can approximate $\mu^2 \sim -m_{H_U}^2$ (for $\tanb\ge 5$) and 
$m_A^2=m_{H_D}^2+m_{H_U}^2 +2\mu^2 \sim m_{H_D}^2-m_{H_U}^2$ at tree level. The
parameters $\mu$ and the sparticle 
spectrum have been computed 
with SuSpect \cite{suspect}. We consider the following collider constraints 
in addition to the bounds on sparticle masses \cite{limits} :

\begin{itemize}
\item {{\bf Higgs boson mass limit:}} 
In the {\it non decoupling} 
region where the $A$ boson
becomes very light so that one has $m_A \sim m_H \sim m_h \sim M_Z$, 
the lower limit of $m_h$ goes down to 93 GeV or even lower. 
We consider that 
the parameter space with $\sin^2(\beta-\alpha)~<~0.3$ (or, 
$\sin(\beta-\alpha)~<~0.55$), but
with Higgs mass $93<m_h<114$ is in agreement with the LEP2 limit \cite{hlim}.
%assumed to belong in the '{\it light Higgs
%boson}' zone. 
In practice, relaxing the constraint marginally we discern this zone with a value
$\sin(\beta-\alpha)~<~0.6$. Consequently, the coupling of the 
heavier Higgs boson to the $Z$ boson
($g_{ZZH} \propto \cos(\beta-\alpha)$)
becomes dominant and this makes the heavier Higgs boson SM - like ($\cos(\beta-\alpha) \sim 1$) 
with the same lower bound as in Eq.(\ref{higgs}). In summary, to obtain 
acceptable SUSY spectra with $93<m_h<114$, in addition to the desired 
value for $\sin(\beta-\alpha)$, one also requires $m_H>114~$GeV.

On the other hand, in the decoupling region 
($\sin(\beta-\alpha) \sim 1$ ), the limit of Eq.(\ref{higgs}) needs to be respected. 
However we note that there is an uncertainty of about 3~GeV
in computing the mass of the light Higgs boson \cite{higgsuncertainty}. This
theoretical uncertainty primarily originates from momentum-independent
as well as momentum-dependent two-loop corrections and higher loop corrections
from the top-stop sector. Consequently, a lower limit of $111$~GeV is often 
accepted for the SUSY light Higgs boson mass.

\item {\bf $Br(b\rightarrow s\gamma)$ constraint:} 
The most significant contributions to $b\rightarrow s\gamma$ 
originate from charged Higgs and chargino exchange diagrams in models like 
mSUGRA.  
The charged Higgs 
($H^-$-$t$ loop) contribution has the same sign and comparable strength 
with respect to the $W^-$-$t$ loop contribution of the SM, which already 
saturates the experimental result. Hence, in scenarios where the charged 
Higgs mass
can be very small, satisfying the $b \rightarrow s \gamma$ constraint requires 
a cancellation between the charged Higgs and the chargino exchange diagrams.
We will choose $A_0$ such that most of the parameter points, especially 
WMAP - compliant ones, can satisfy this constraint.   
The effect
of this constraint in the {\it light Higgs boson} zone has been discussed
in \cite{kim}.  We have 
used the following
 3$\sigma$ level constraint from $b \rightarrow s \gamma $ with 
the following limits \cite{bsg-recent}. 
\begin{equation}
2.77 \times 10^{-4} < Br (b \rightarrow s \gamma) < 4.33 \times 10^{-4}.
\label{bsgammalimits}
\end{equation}

\item {\bf $Br(B_s\rightarrow \mu^+ \mu^-)$ constraint:} 
Similarly, the flavor physics observable 
$B_s\rightarrow \mu^+ \mu^-$ may become very significant in 
this particular parameter space. The current experimental limit 
for $Br(B_s \to \mu^+ \mu^-)$ 
coming from CDF \cite{CDF} can be written as (at ${\rm 95\,\%\,C.L.}$) 
\begin{eqnarray}
{\rm Br} ( B_s \to \mu^+ \mu^-) < 5.8 \times 10^{-8}, 
\label{Bsmumu}
\end{eqnarray}
which has recently been improved to $< 4.3 \times 10^{-8}$ 
at $95\%$ C.L \cite{Morello:2009wp}.
The estimate of 
$ B_s \to \mu^+ \mu^- $ in the MSSM depends strongly
on the mass of A-boson and on the value of $\tan \beta$. In particular, the  
neutral Higgs boson contribution scales as $ m_A^{-4}$ whereas 
there is a $(\tan \beta)^6$ type of dependence. However, in the
present analysis, we choose $\tan\beta~(=10)$ which makes this constraint
less restrictive for most of the parameter space.

\item {\bf WMAP constraint :}
In computing the relic density constraint, we consider 
the following 3$\sigma$ limit of the WMAP data \cite{WMAPdata}, 
\begin{equation}
0.091 < \Omega_{CDM}h^2 < 0.128.
\label{relicdensity}
\end{equation}
Here $\Omega_{CDM}h^2$ is the dark matter 
relic density in units of the critical
density and $h=0.71\pm0.026$ is the Hubble constant in units of
$100 \ \rm Km \ \rm s^{-1}\ \rm Mpc^{-1}$. We use the 
code micrOMEGAS \cite{micromegas} to 
compute the neutralino relic density.

\end{itemize}

%%%%%%%%%%%%%%%%%%%%%%%%%%%%%%%%%%%%%%%%%%%%%%%%%%%%%%%%%%%%%%%%%%%%%%%%%%%%%%%%
\section{Direct Detection of Dark Matter}\label{dir_detection}
Direct detection of the
LSPs involves measurement of recoil energies of nuclei as they are scattered
by WIMPs.
The effective Lagrangian that describes $\neut1-q$ 
elastic scattering at
small velocities is given by (for reviews see \cite{dmreview})
\begin{equation}
{\cal L} = \alpha'_{qi}\bar{\neut1} \gamma^\mu \gamma^5 \neut1 \bar{q_{i}}
\gamma_{\mu} \gamma^{5} q_{i} +
\alpha_{qi} \bar{\neut1} \neut1 \bar{q_{i}} q_{i}~.
\label{lagxsection}
\end{equation}
The first term represents spin-dependent scattering
while the second term refers to spin-independent scattering.
Eq.(\ref{lagxsection}) assumes summing over both the quark
generations $q$ while the subscript $i$ runs for up ($i=1$) 
and down type ($i=2$) quarks respectively.
The neutralino-quark coupling coefficients
$\alpha_{q}$ and $\alpha'_{q}$ contain all SUSY model-dependent 
information.
The spin-independent scattering cross-section of a neutralino with a target
nucleus of proton number (atomic number) 
$Z$ and neutron number $A-Z$ ($A$ being the
mass number) is given by
\begin{equation}
\sigma^{SI} = \frac{4 m_{r}^{2}}{\pi} \left[ Z f_{p} + (A-Z) f_{n}
\right]^{2}~.
\label{sitotal}
\end{equation}

\noindent
Where $m_r$ is the reduced mass
defined by $m_r=\frac{m_{\neut1} m_N}{(m_{\neut1}+ m_N)}$ and
$m_N$ refers to the mass of the nucleus.
The quantities $f_p$ and $f_n$
contain all the information of short-distance physics and nuclear
partonic strengths. These are given by
\begin{equation}
\frac{f_{p, (n)}}{m_{p, (n)}} = \sum_{q=u, d, s} f_{Tq}^{(p, (n))} 
\frac{\alpha_{q}}{m_{q}} +
\frac{2}{27} f_{TG}^{(p, (n))} \sum_{c, b, t} \frac{\alpha_{q}}{m_q}~, 
\label{fpn}
\end{equation}
where $f_{Tq}^{(p, (n))}$ defined as
\begin{equation}
m_{p, (n)} f_{Tq}^{(p, (n))} = \langle p, (n) | m_{q} \bar{q} q | p, (n) 
\rangle \equiv 
m_q B_q~. 
\label{defbq}
\end{equation}
The quantities $f_{Tq}^{(p, (n))}$ can be evaluated using
a few hadronic data \cite{ellis2000}.
The gluon - related part namely $f_{TG}^{(p, (n))}$ is given 
by

\begin{equation}
 f_{TG}^{(p, (n))} = 1 - \sum_{q=u, d, s} f_{Tq}^{(p, (n))}~. 
\end{equation}
The numerical values of $f_{Tq}^{(p, (n))}$ 
may be seen in \cite{ellis2000,hoopertasi}. 
We compute the neutralino-nucleon spin-independent scattering cross-section
by using the code DarkSusy \cite{darksusy}. 
We should note here that the parameter
$f_{Ts}^{(p, (n))}$ requires the information of 
the pion-nucleon sigma term $\sigma
_{\pi N}$ and the size of the $SU(3)$ symmetry breaking-$\sigma_0$ as
$f_{Ts}^{(p, (n))} \propto (\sigma_{\pi N}-\sigma_0)$,
so that the leading contribution in $\sigma^{SI}$ goes as  $\sim (\sigma_{\pi N}-\sigma_0)^2$. 
Clearly, uncertainty in the computation of $\sigma_{\pi N}$ 
%(thus in $y$) 
can produce significant
shift to the strange quark parameter $f_{Ts}^{(p, (n))}$ and consequently to
the measured spin-independent cross-section.
Recent lattice results hint towards
much smaller values of the
$f_{Ts}^{(p, (n))}\sim 0.02$ \cite {Cao:2010ph,lattice}, 
a value much smaller than previous estimates. Considering even larger
uncertainty in $\sigma_{\pi N}$ 
one may assume 
$\sigma_{\pi N}=\sigma_0$ 
which leads to 
$f_{Ts}^{(p, (n))} = 0$ \cite{Ellis:2008hf}.
In DarkSusy \cite{darksusy} the above
coefficient is chosen as $f_{Ts}^{(p, (n))} \equiv 0.14$.   
This could provide a significant
change in the results of the $\sigma^{SI}$. 
In fact, in 
 \cite{Cao:2010ph,Ellis:2008hf}, the
variation in the spin-independent cross-section due to this reduced $f_{Ts}^{(p, (n))}$
has been estimated. We shall also
compute
the variation in the spin-independent cross-section
with the reduced values of $f_{Ts}^{(p, (n))}$. Similarly, we shall
comment on the other sources
of uncertainties and consequently their effects on $\sigma^{SI}$
and the corresponding experimental limits.

The scalar cross-section depends on t-channel Higgs exchange
($h,H$) and s-channel squark exchange diagrams 
($\sigma^{SI} \sim \frac{1}{m^4_{H,h,\tilde q}}$). Now, considering the
strong bounds on the light squark masses, 
$\sigma^{SI}$ is dominated by the exchange
of CP-even Higgs bosons.
In the present context, where
both CP-even Higgs masses are quite small, $\sigma^{SI}$ becomes enhanced. 
In fact, we shall see that in the mass range 
$93<m_h<114$, the spin-independent
cross-section becomes maximal and this renders the 
{\it light Higgs boson}
 zone highly constrained even for $\tanb=10$.    
Apart from the masses of the Higgs
bosons, the cross-section depends strongly on the couplings  
$\neut1 \neut1 h(H)$ (${\cal{C}}_{\neut1 \neut1 h(H)}$) 
and also on $h(H) q \bar q$ (${\cal C}_{q \bar q h(H)}$). 
The latter contributions in the
context of down - type fermions include $\tanb$ as 
${\cal C}_{q \bar q h} \sim \tan\beta \cos(\beta-\alpha)$ and
${\cal C}_{q \bar q H} \sim \tan\beta \sin(\beta-\alpha)$.
Clearly, both couplings may become quite large for larger values of $\tanb$.
These large couplings, in addition to the small mass values for $m_{h,H}$,
make this {\it light Higgs boson} zone almost ruled out for larger values
of $\tanb$.

\section{Indirect Detection of Dark Matter}\label{ind_detection}
Indirect detection techniques are based on the detection of 
primary or secondary particles resulting from dark matter (DM) 
annihilation or decay.
The observed flux of these particles is proportional to the annihilation rate of the
DM species which in turn is, for the annihilating case, proportional to  $\rho^2$, the
dark matter density profile. 
In the following, we summarize the basic ingredients of the formalism that 
we shall
be employing in this work. 
\subsection{Gamma-ray Detection}
\subsubsection{The flux at intermediate latitudes}
\label{gammarays}
The differential flux of gamma--rays generated from dark matter annihilations
and coming from a direction forming an angle $\psi$ with respect to
the galactic center (GC) is
\begin{equation}
 \frac{d\Phi}{dE}(\psi, E) = N \frac{1}{4 \pi} 
\frac{\left\langle \sigma v \right\rangle }{m_{\chi}^2} 
\sum_{f}\frac{dN_f}{dE} 
\int_{los} \rho^2(l(\psi)) dl(\psi),
\end{equation}
where $N = 1, 1/2$ if the annihilating particle is Majorana-like or Dirac-like
respectively, $\left\langle \sigma v \right\rangle$ is the total self annihilation
cross-section averaged over velocity for $v \rightarrow 0$, $m_\chi$ is the mass of any DM particle
, $dN_f/dE$ is the differential spectrum of final state $f$ into gamma-rays
of energy $E$ and $\rho$ is the dark matter density profile. We note that in this
work we only focus on prompt photons, ignoring potential contributions from 
synchrotron emission or inverse Compton scattering of charged particles with the
interstellar medium.

The functions $\frac{dN_f}{dE}$  have been computed using the fragmentation of Standard Model 
particles into $\gamma$-rays according to the PYTHIA \cite{Sjostrand:2006za} Monte Carlo code. 

Now, of course, we don't observe annihilations along a line but rather within a conical 
region around some angle $\psi_0$. It is convenient to define the dimensionless quantity
\begin{equation}
 J(\psi) = \frac{1}{R_0}\frac{1}{\rho_0^2} \int_{los}\rho^2(l(\psi)) dl(\psi),
\label{Jquantity}
\end{equation}
where $R_0$ is the sun's distance from the galactic center and $\rho_0$ the local
DM density (actually, they can be any arbitrary normalization factors rendering $J$ 
a dimensionless quantity).\\
Then, we can define the average of this quantity within a solid angle $\Delta\Omega$
as
\begin{equation}
 \bar{J}(\Delta\Omega, \psi_0) = \frac{1}{\Delta\Omega} \int_{\Delta\Omega(\psi_0)} J(\psi)d\Omega,
\label{Jbar}
\end{equation}
and we can then rewrite the flux as
\begin{equation}
  \frac{d\Phi}{dE}(\psi, E) = \frac{N}{4 \pi} \frac{\left\langle \sigma v \right\rangle }{m_{\chi}^2} \left( \sum_{f}\frac{dN_f}{dE} \right) R_{0} \rho_{0}^2 \ \bar{J}(\Delta\Omega, \psi_0).
\label{calcflux}
\end{equation}

The most common region of the sky that is examined in the literature as a source of $\gamma$-rays
is the galactic center, since it is the region where $N$-body simulations predict a maximization
of the dark matter density distribution and, hence, the corresponding gamma-ray flux.
However, the galactic center is a region which is known poorly
: There are large uncertainties in the background modelizations as well
as the density profile itsself. 

As an alternative, it has been proposed (see, for example, ref. \cite{Stoehr:2003hf}) that one could maximize the
signal/background ratio by actually excluding the region around the galactic center. Following
this reference, we perform our computations in a conoidal region extending from $20^\circ$ up to 
$35^\circ$ from the galactic center, excluding at the same time the regions within $10^\circ$ from the
galactic plane. It has actually been shown that within the framework of such an analysis, 
one can enhance the signal/background ratio by up to roughly an order of magnitude. The relevant
background for this region will be discussed in the next subsection. It should also be noted that
other regions of the sky might provide very interesting results, as shown for example in \cite{Cholis:2009gv}.

In the meantime, we present in table \ref{Jbars} the values obtained for the $\bar{J}$ quantity
defined in eq.(\ref{Jbar}), for three different halo profiles often discussed in the literature:
The Navarro, Frenck and White one (NFW) \cite{Navarro:1995iw}, which seems to be favoured by the recent results of the 
Via Lactea II simulation \cite{Diemand:2008in}, the Einasto profile which is  
 favoured by the findings of
the Acquarius Project collaboration \cite{Navarro:2008kc} and a NFW-like profile that has tried to take into account
the effects of baryons in the inner galactic regions, causing an adiabatic collapse of DM
in this area and a significant enhancement of its central density \cite{Prada:2004pi}.

\begin{table}
\begin{center}
\begin{tabular}{|c|ccccc|}
\hline
 & $a$ [kpc] & $\alpha$ & $\beta$ & $\gamma$ & $\bar{J}$\\
\hline
Einasto &  -   &   -   &       &   -    & $ 10.486$\\
NFW     & $20$ & $1.0$ & $3.0$ & $1.0$  & $8.638$\\
NFW$_c$ & $20$ & $0.8$ & $2.7$ & $1.45$ & $ 12.880$\\
\hline
\end{tabular}
\caption{{\footnotesize Einasto, NFW and NFW$_c$
density profiles with the corresponding parameters,
and values of $\bar{J}(\Delta\Omega)$ for the galactic
region under consideration.}}
\label{Jbars}
\end{center}
\end{table}
The $\bar{J}$ values obtained in the table demonstrate another virtue of searching
for dark matter at intermediate latitudes, namely the fact that the results become
quite robust with respect to the various dark matter density distribution
modelizations. In the case of the galactic center, there can be differences of
orders of magnitude in this factor, whereas in this case the differences are of
$O(1)$. In the following, we shall be presenting our results for an Einasto
profile, since it yields results somewhere in the middle among the other two
scenarios.

\subsubsection{The Fermi 1-year observations}
The Fermi collaboration has published its 1-year observation results outside the GC
 \cite{Abdo:2010nz}. In this paper, the collaboration presents its
observations for a period of $19$Msec and for various galactic latitudes. 
In the companion paper, the results for latitudes $20^\circ<b<60^\circ$ are
also presented, which lie actually in our region of interest. The data between $b = 10^\circ$ and
$20^\circ$ are included in the same paper, presenting an enhancement by a factor of roughly 
$1.5 - 2$  with respect to the higher latitude data. In this analysis, 
for the sake of simplicity, we shall
be focusing on the data from higher latitudes ($20^\circ<b<60^\circ$), integrating them over the whole
region of interest. This is  justified, since we have excluded 
from our analysis the region within $20^\circ$ from the galactic center, which should provide
one of the major contributions to this spectrum.

In the paper, the authors could fit the data quite well using a Diffuse Galactic
Emission model based on the GALPROP code. We find this model to be well reproduced, in our 
region of interest, by a simple power-law
\begin{equation}
 \Phi_{\mbox{\begin{tiny}bkg\end{tiny}}}^{\mbox{\begin{tiny}Th\end{tiny}}} = 2.757\cdot 10^{-6} E^{-2.49}
\end{equation}
in units of GeV$^{-1}$ sec$^{-1}$ cm$^{-2}$ sr$^{-1}$.
In the same analysis, the collaboration presents the detector effective area values 
that should be used in order to compare predictions with observations, as a function of
the gamma-ray energy. In the following, we shall be using these values rather than the usual
nominal effective area of $10000$ cm$^2$.

%%%%%%%%%%%%%%%%%%%%%%%%%%%%%%%%%%%%%%%%%%%%%%%%%%%%%%%%%%%%%%%%%%%%%%%%%%%%%%%%
\subsection{Antiproton Detection}\label{antiprotons}
Here, the charged particles present
the complication of propagating throughout the Interstellar Medium (ISM) compared to
the gamma-ray detection. Efforts have been made to describe the physics behind cosmic-ray propagation. Different
treatments make different sets of assumptions and utilise different formalisms under specific
assumptions  \cite{Baltz:1998xv, Strong:1998su, Lavalle:2006vb,Maurin:2002ua,Maurin:2006hy,Maurin:2006ps,Lavalle:1900wn}.
Such treatments have of course been applied in the case of the MSSM neutralino dark matter detection
 \cite{Mambrini:2006aq, Baltz:2001ir, deBoer:2002nn, Hooper:2008kv}.

In this work, we shall use the two-zone diffusion model and its semi-analytical 
solution as described, for example, in references \cite{Maurin:2006hy,Lavalle:1900wn}.
In this model, positron and antiproton propagation takes place in a cylindrical region
(Diffusive Zone, DZ) around the galactic center of half thickness $L$.
Cosmic rays can escape this region, a case in which they are simply lost.
\\
The master equation for cosmic-ray propagation is a diffusion-convection-reacceleration equation:
\begin{equation}
\partial_t\psi+ \partial_z(V_c\,\psi) - \nabla(K\,\nabla\psi) - \partial_E \left[ b(E)\,\psi + K_{EE}(E)\,\partial_E\,\psi\right] = q\,,
\label{masterProp}
\end{equation}
where $\psi = dn/dE$ is the space-energy density of the positrons or antiprotons, 
$b(E)$ is the energy loss rate,
$q$ is the source term,  $V_c \approx (5 - 15)$ km/s is the convective wind velocity wiping
away the positrons or antiprotons from the galactic plane, 
\begin{equation}
K(E) = K_0\,\beta
\left( \frac{E}{E_0} \right)^\alpha
\label{DiffCoeff}
\end{equation}
is the diffusion coefficient, with $\beta$ being the particle's velocity, $K_0$ the diffusion
constant, $\alpha$ a 
constant slope, $E$ the kinetic energy (for positrons in practice the total
one), $E_0$ a reference energy (which we take to be $1$ GeV), and
\begin{equation}
K_{EE} = \frac29\,V_{a}^{2}\,\frac{E^2\,\beta ^4}{K(E)}
\end{equation}
is a coefficient describing reacceleration processes.\\

%%%%%%%%%%%%%%%%%%%%%%%%%%%%%%%%%%%%%%%%%%%%%%%%%%%%%%%%%%%%%%%%%%%%%%%%%%%%%%%%%%%%%%%%%%%%%%%%%%%%%%%
\subsubsection{Differential Event Rate}
\label{AntiprotonFlux}
In antiproton propagation, all energy redistributions in the initial 
(injection) spectrum --energy losses, reacceleration, 
as well as `tertiary' contributions (i.e., contributions
from antiprotons produced upon inelastic scattering of other antiprotons with the interstellar 
medium) can be ignored. 
%Whether these redistributions are important 
%or not depends mainly on the antiproton energy. 
The importance of these redistributions depends on the antiproton energy.
For GeV energies, 
the results may deviate up to $50\%$ from those obtained with the  
(more complete) Bessel function treatment\footnote{In reference \cite{Maurin:2006hy}
a comparison between the two methods can be found (see figure $2$).}.
But for energies around $~10$ GeV, the accuracy
of the method improves dramatically, yielding essentially indistinguishable
results at slightly higher energies. Considering that the $\bar{p}$ energy region
begins at $10$ GeV, we can safely use this simplified approach. 

Let us denote $\Gamma_{\overline{p}}^{\mbox{\tiny{ann}}} = \sum_{\mbox{\tiny{ISM}}} 
n_{\mbox{\tiny{ISM}}}\,v\,\sigma_{\overline{p} \ \mbox{\tiny{ISM}}}^{\mbox{\tiny{ann}}}$,
the destruction rate of antiprotons in the interstellar medium, where $\mbox{ISM} = \mbox{H and He}$, 
$n_{\mbox{\tiny{ISM}}}$ is the average number density of ISM in the galactic disk, $v$ is the
antiproton velocity and $\sigma_{\overline{p} \ \mbox{\tiny{ISM}}}^{\mbox{\tiny{ann}}}$ is the
$\bar{p} - \mbox{ISM}$ annihilation cross-section. 
Implementing the aforementioned simplifications, 
the transport 
equation for a point source (which actually defines the propagator $G$) is:
\begin{equation}
\left[ -K\,\nabla + V_c\,\frac{\partial}{\partial z}
+2\,h\,\Gamma^{\mbox{ann}}_{\bar p}\,\delta(z) \right] G = 
\delta\left(\vec{r} - \vec{r'}\right)\,,
\label{PropagatorDef}
\end{equation}
with $h = 100$ pc being the half-thickness of the galactic disc. In this equation, the origin of the 
coordinate system is taken to be the Galactic Center, whereas $r = |\vec{r}_{\odot} - \vec{r}|$
is the distance of a given point from the sun. The presence of the $\delta$ function in the LHS
of eq.\eqref{PropagatorDef} reflects the fact that we work in the so-called ``thin disk'' approximation, 
where the size of the diffusive zone is considered to be much larger than the size where spallations
can take place \cite{Donato:2001ms}.
The antiproton propagator connecting the solar position and any point in the diffusive zone 
can then be written (in cylindrical coordinates) as
\begin{equation}
G^{\odot}_{\overline{p}}(r,z) = 
\frac{e^{-k_v\,z}}{2 \pi\,K\,L}\,
\sum_{n=0}^{\infty} c_n^{-1}\,K_0\left(r\sqrt{k_n^2 + k_v^2}\right)
\sin(k_n\,L)\,\sin(k_n\,(L-z))\,,
\label{GreenPbars}
\end{equation}
where
$K_0$ is a modified Bessel function of the second kind and
\begin{eqnarray}
c_n & = & 1 - \frac{\sin(k_n L) \cos(k_n L)}{k_n L}\,,\\
k_v & = & V_c/(2K)\,,\\
k_d & = & 2\,h\,\Gamma_{\overline{p}}^{\mbox{\tiny{ann}}}/K + 2\,k_v\,.
\end{eqnarray}
$k_n$ is obtained as the solution of the equation
\begin{equation}
n\,\pi - k_{n}\,L - \arctan(2\,k_n/k_d) = 0, \ \ n\in\mathbb{N}\,.
\end{equation}
Then, in order to compute the flux expected on  earth, we should
convolute the Green function \eqref{GreenPbars} with the source 
distribution $q(\vec{x}, E)$. For the dark matter annihilations in the 
galactic halo, the source term is given by
\begin{equation}
q(\vec{x}, E) = \frac{1}{2}
\left( \frac{\rho(\vec{x})}{m_\chi} \right)^2
\sum_i 
\left( 
\langle\sigma v\rangle \frac{dN_{\bar{p}}^i}{dE_{\bar{p}}}
\right)\,,
\label{eq:q}
\end{equation}
where the index $i$ runs
over all possible annihilation final states. The decay and hadronization of SM 
particles into antiprotons are calculated with {\tt PYTHIA} \cite{Sjostrand:2006za}.  
Now the distribution of dark matter in the Galaxy $\rho(\vec{x})$, 
is not the dominant factor in the calculation
of the antiproton flux, thus we assume a standard NFW profile for the sake of definitiveness.
\\
The antiproton flux on the earth can finally be expressed as
\begin{equation}
\Phi_{\odot}^{\bar{p}} (E_{\mbox{\tiny{kin}}}) = 
\frac{c\,\beta }{4\pi}
\frac{\langle\sigma v\rangle}{2}
\left(   \frac{\rho(\vec{x}_{\odot})}{m_\chi} \right)^2
\frac{dN}{dE}(E_{\mbox{\tiny{kin}}})
\int_{DZ} \left(\frac{\rho(\vec{x_s})}{\rho(\vec{x}_{\odot})} \right)^2
G^{\odot}_{\overline{p}}(\vec{x}_s)\,d^3x_s\,.
\label{PbarFlux}
\end{equation}
Notice that since we ignore energy redistributions, the production and detection energy are the
same. 
We hence compute the integral in equation \eqref{PbarFlux} only once for each value of the injection energy
(which is the same as the detection energy) by means of a VEGAS Monte-Carlo algorithm
and use its values throughout our parameter space scan.

Regarding  the propagation parameters $L$, $K_0$, $\alpha$, and  $V_c$, we take their values from 
the well-established MIN, MAX and MED models --see table \ref{PropParameters}. 
The MIN and MAX models correspond to the minimal and maximal antiproton
fluxes that are compatible with the B/C data. 
The MED model, on the other hand, corresponds to  the parameters that best fit the B/C data.
\begin{center}
\begin{table}
\centering
\begin{tabular}{|c|cccc|}
\hline 
&$L$ [kpc]&$K_0$ [kpc$^2$/Myr]&$\alpha$&$V_c$ [km/s]\\
\hline 
MIN & $1$ & $0.0016$ & $0.85$ & $13.5$\\ 
MED & $4$ & $0.0112$ & $0.70$ & $12.0$\\
MAX & $15$ & $0.0765$ & $0.46$ & $5.0$\\
\hline 
\end{tabular}
\caption{{\footnotesize Values of propagation parameters
widely used in the literature and that provide minimal and maximal antiproton fluxes,
or constitute the best fit to the B/C data.}}
\label{PropParameters}
\end{table}
\end{center}

%%%%%%%%%%%%%%%%%%%%%%%%%%%%%%%%%%%%%%%%%%%%%%%%%%%%%%%%%%%%%%%%%%%%%%%%%%%%%%%%%%%%%%%%%%%%%%%%%%%%%%%
\subsubsection{The background and antiproton detection with AMS-02}
AMS-02 \cite{Goy:2006pw} is expected to be able to measure antiproton fluxes 
with an average geometrical acceptance of $330$ cm$^2$ sr and energy above $16$ GeV and up to 
$300$ GeV. 
As mentioned in paragraph \ref{AntiprotonFlux}, we stick to 
energies above $10$ GeV in this analysis.
Similarly, we consider a $3$-year run and the mentioned energy range is divided 
into $20$ logarithmically evenly spaced energy bins.

At the same time, the PAMELA collaboration have recently published its 
updated antiproton measurements
in the kinetic energy range from $60$ MeV up to $180$ GeV \cite{:2010rc}. 
The data acquisition period
was $850$ days and the results seem to be in good agreement with 
several theoretical
predictions for secondary production. 
Model-independent
constraints from this data have been discussed, for example, in \cite{Cholis:2010xb}.

Above $10$ GeV, which is the region of interest in our case, the data can be well described by a simple
power law
\begin{equation}
 \Phi_{\mbox{\begin{tiny} bkg \end{tiny}}} = 5.323\times 10^{-4} E^{-2.935}
\ \ \mbox{GeV}^{-1} \mbox{sec}^{-1} \mbox{sr}^{-1} \mbox{cm}^{-2}.
\end{equation}

%%%%%%%%%%%%%%%%%%%%%%%%%%%%%%%%%%%%%%%%%%%%%%%%%%%%%%%%%%%%%%%%%%%%%%%%%%%%%%%

\section{Results and Discussion}
\label{resultsall}

\subsection{Cold Dark Matter Constraint and Light Higgs Boson Scenario} 
\begin{figure}[ht!]
\begin{center}
\includegraphics[width=12cm,height = 10cm]{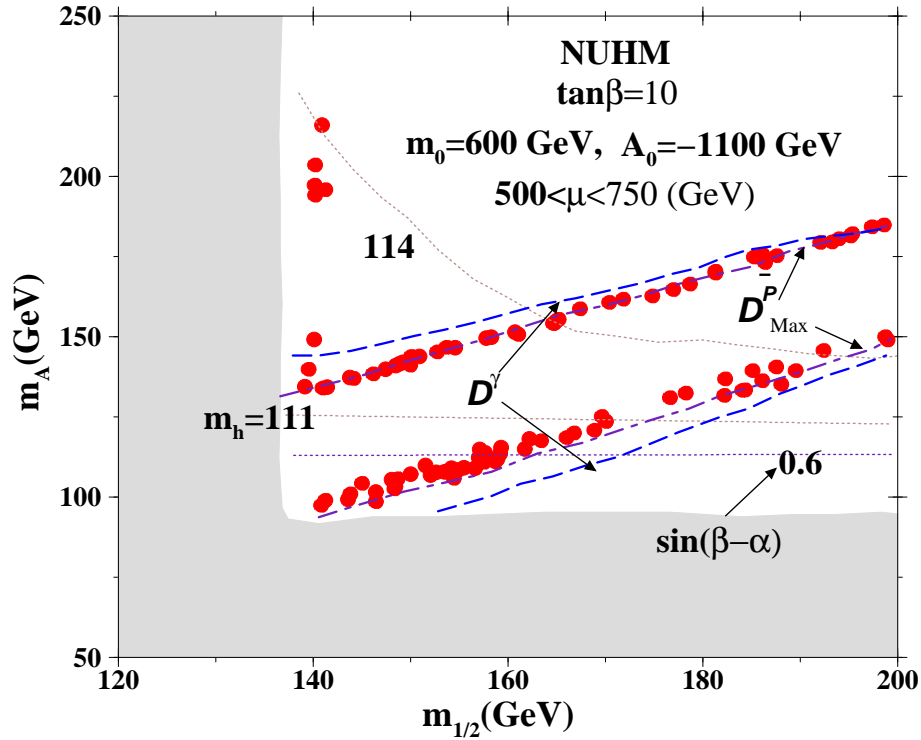}\\
\end{center}
\caption{WMAP satisfied parameter points in the $\mhalf-m_A$ plane.
 Acceptable relic density is obtained via $s-$channel $h$,  $A$ or $H$ exchange. 
 Neutralino masses of $\sim 55-65$ GeV 
correspond to the {\it light Higgs boson} region. Detectability of the
photon and anti-proton signal are represented by $D^{\gamma}$ and $D^{\bar p}_{Max}$. The WMAP - compliant $A$ pole annihilation region is within the reach of Fermi-GLAST and upcoming AMS-02 experiments.}
\label{a01100}
\end{figure}

\begin{figure}[ht!]
\begin{center}
\includegraphics[width=12cm,height = 10cm]{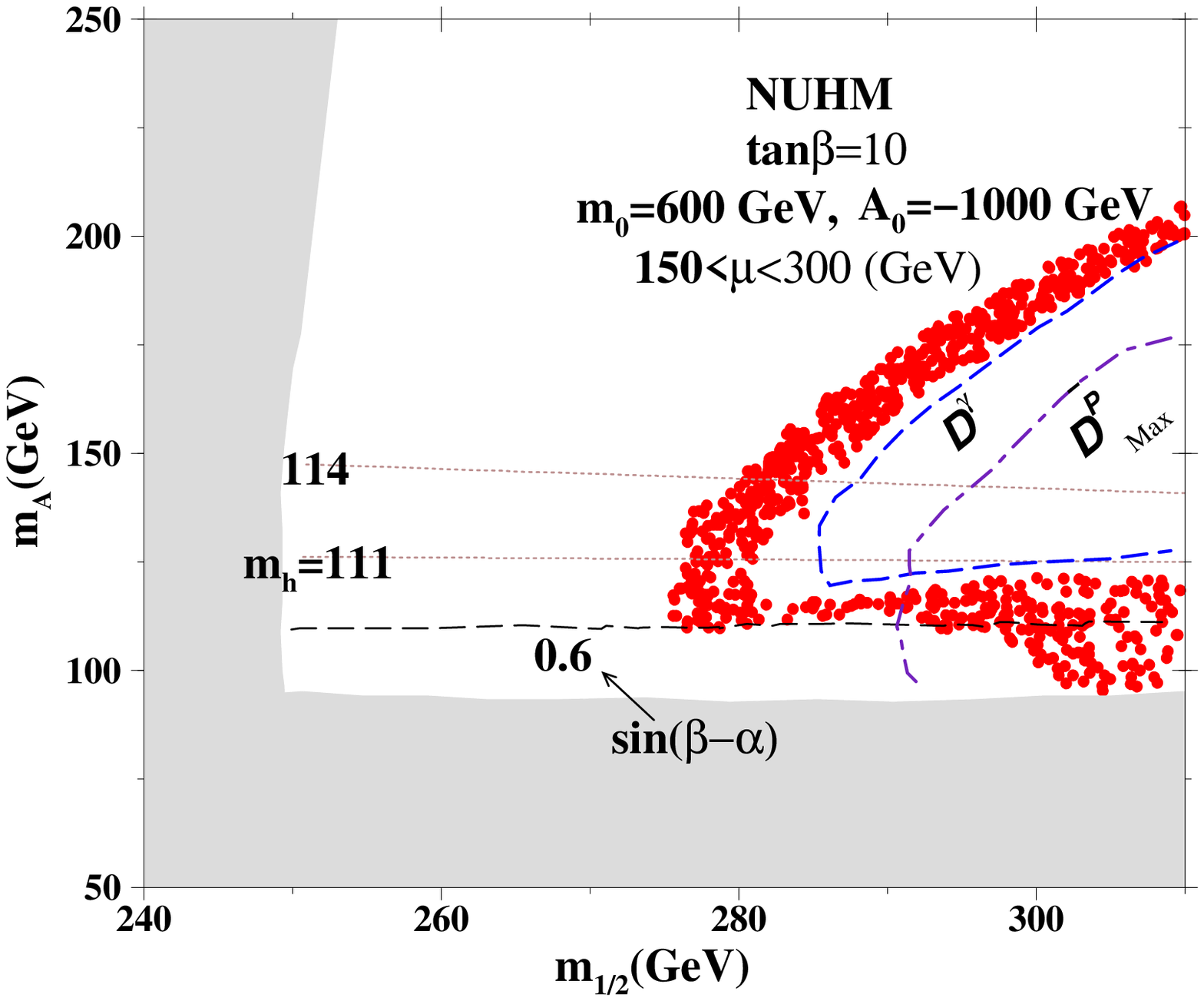}\\
\end{center}
\caption{Same as Fig.\ref{a01100}, except {\it light Higgs boson} zone
is shifted to the larger values for neutralino masses. 
}
\label{a01000}
\end{figure}

We present our main results in Figs.\ref{a01100} and \ref{a01000}, where we depict the valid
parameter space consistent with the WMAP constraint in the $\mhalf-m_A$ plane, 
while assuming a moderate
value for $\tanb(=10)$. Let us first concentrate on Fig.\ref{a01100}. 
In this case, the other parameters are (i) $m_0=600~$GeV, (ii) $A_0=-1100$
GeV{\footnote {We consider a
top pole mass $m_t = 173.1$~GeV \cite{cdfd0}. Similarly we choose $sign(\mu)>0$ in this analysis.}}. We have 
varied the mass parameters $\mhu$ ($0<\mhu<m_0^2$) and $\mhd$ 
($-1.5 m_0^2<\mhd<-0.5m_0^2$) to obtain light neutralino dark matter consistent with light Higgs masses ($m_{H,A}\le 250~$GeV) at the electroweak scale.
As mentioned in 
Sec.\ref{sec:model}, $\mu$ and $m_A$ are the derived quantities which we have 
calculated using Eqs.\eqref{rewsbmusqeqn} and \eqref{rewsbBeqn}.

%the $\mu$ and $m_A$ parameter in the
%desired range. 
%within $0<\mhu<m_0^2$ and $-1.5 m_0^2<\mhd<-0.5 m_0^2$. 
%This provides us with
%reasonably larger values of the $\mu$ parameter for the valid parameter points.
We are particularly concerned about the {\it light Higgs boson} zones and
this, in the present case, requires $\mhalf$ values $\le 160~$GeV.
The lightest neutralino in the form of $\tilde B$ (with a non negligible $\tilde H$ component) 
produces the acceptable relic density and we represent this through red circles.
We also show contours for the lightest Higgs mass ($m_h$) and
$\sin(\beta-\alpha)$. All points in the parameter space with $\sin(\beta-\alpha)< 0.6$
are considered to belong to the {\it light Higgs boson} zone where the Higgs mass can
 evade the LEP2 limit due to
the reduced coupling with the $Z$ boson. On the other hand, admitting 
a 3 GeV uncertainty in the Higgs mass calculation, as mentioned, we delineate
the regions corresponding to $111$ GeV $< m_h < 114$ GeV. With a smaller $\mhalf$ value, 
the WMAP - allowed region predicts lighter squarks as well as lighter gluinos.
\\
There are two distinct regions in the parameter space satisfying the 
relic abundance constraint:

$(a)$ The light Higgs pole annihilation region where
 neutralino annihilation 
produces an acceptable relic density via
the s-channel exchange of a light Higgs boson. This particular
region extends in the direction of $m_A$ with gaugino mass value $\sim 140~$GeV.
This zone, however, is highly constrained in the mSUGRA model if one respects
the flavor physics constraints \cite{djouadiLightHiggs,myhiggs}. 
The spin-independent cross-sections \cite{myhiggs},
on the other hand, could be compatible with the CDMS-II \cite{cdms-2} limits.

($b$) The second region is the funnel region where the
annihilations are principally due to exchange of $A$ and $H$ bosons, 
with $2m_{\chi_1^0} \simeq m_A, m_H$. 
Like in mSUGRA, 
this WMAP - satisfying
region in the NUHM is principally characterized by the pseudoscalar 
Higgs boson -  mediated resonant annihilation.  The exact or near-exact 
resonance regions have very large annihilation cross-sections resulting in  
a high degree of under-abundance of dark matter. In fact, acceptable relic density
can be produced when the $A$-width is 
quite large and $2m_{\chi_1^0}$ can be appreciably away from the exact 
resonance zone. This is precisely the reason for the two branches of red circles 
that extend along the direction of $\mhalf$ in Fig.\ref{a01100}.

As already mentioned in the introduction, our primary interest 
is related to the region where $\sin(\beta-\alpha)< 0.6$. A lightest
neutralino with mass $55<m_{\chi^0_1}<65$ GeV falls in this particular
category with the $A$ boson playing the dominant role in the annihilation process. 
Now, apart from the mass of the $A$ bosons, 
neutralino pair annihilation also depends on the coupling 
${\cal{C}}_{\chi_1^0 \chi_1^0 A}$ which goes as the product of the Bino and Higgsino components
of the LSP  ($N_{11}$ and $N_{13},
N_{14}$ respectively). 
%The coupling depends on both the 
%electroweak gaugino and Higgsino composition of the
%neutralino as ${\cal{C}}_{\chi_1^0-\chi_1^0-A(H)} \sim 
%(g_2 N_{12}-g_1 N_{11})(N_{14}cos\beta - N_{13}sin\beta)$. 
%With a small
%Higgsino component, the Bino-dominated LSP may undergo resonant annihilation 
%mechanism via the $S$ channel pole to provide the acceptable relic density.
We recall that the Higgsino components of the LSP
are essentially determined by the $\mu$ parameter. For a relatively large $\mu$
 (500~GeV$~<\mu<~$750 GeV, Fig.\ref{a01100}), the Higgsino component is relatively small 
which essentially diminishes the coupling ${\cal{C}}_{\chi_1^0 \chi_1^0 A}$.
In this regime, one finds that
light neutralinos with mass ($m_{\neut1} \sim 55-65$ GeV)
can satisfy the WMAP constraint in the {\it non-decoupling}
zone (i.e. where $m_A\sim100 $ GeV). We shall also present a scenario where a reasonably
heavier neutralino ($m_{\neut1} \sim 100$ GeV) can pair-annihilate efficiently
in the {\it non-decoupling} zone via the enhancement of the ${\cal{C}}_{\chi_1^0 \chi_1^0 A}$ coupling.
Now, as neutralino masses increase, the funnel region extends in the direction of larger $m_A$
and the Higgs bosons fall into the decoupling
region (see Fig.\ref{a01100}). 
Interestingly, even this WMAP - satisfying zone does not require
large values of gaugino masses. We will observe that this whole region
can lead to large gamma-ray as well as antiproton signals in present or oncoming
experiments. 

Before presenting our results on indirect detection, we would like
to point out the results of the flavor physics observables like $b \rightarrow
s \gamma$ and $B_s \rightarrow \mu^+ \mu^-$. Since we have chosen a rather 
moderate value for $\tanb$ ($=10$), the $B_s \rightarrow \mu^+ \mu^-$ constraint is 
not quite stringent for the parameter space shown in Fig.\ref{a01100}.
On the other hand $Br(b \rightarrow s \gamma)$ constitutes a strong
constraint particularly in the {\it non-decoupling} region 
where charged Higgs bosons can be very light. However, 
 with large negative $A_0$ values (negative values for $A_t$ at the electro-weak
scale), one of the stop eigenstates
becomes lighter due to large mixing and this provides a cancellation
between charged Higgs and chargino induced diagrams. Choosing $A_0=-1100~$GeV 
at $M_{GUT}$, almost all parameter points and more importantly the whole WMAP
allowed region in the $\mhalf-m_A$ plane can comply with the constraint. 
Now, in regard to the gray regions (Fig.\ref{a01100}) one observes that:
 (i) For $\mhalf \ge 135~$GeV parameter space points with
 $m_A$ smaller than roughly $100~$GeV 
are not compatible with the Higgs mass limit in the {\it non-decoupling} 
zone i.e.,
here one has $m_h<93~$GeV, (ii) for $\mhalf \le 135~$GeV, the gluino becomes 
lighter (we impose $m_{\tilde g}> 390~$GeV for a valid parameter 
space point \cite{Amsler:2008zzb}) and
then very soon the chargino becomes too light with $m_{\chi_1^\pm}<103.5~$GeV. We should
note here that a light Higgs  with mass $m_h\le 93~$GeV may be allowed,
but then $\sin(\beta-\alpha)$ needs to be further suppressed. This
region is then further constrained and we did not consider it in our analysis.

As discussed, with the choice of parameters of Fig.\ref{a01100}, 
the {\it light Higgs boson} region
is confined only up to $m_{\chi_1^0}\sim 65~$GeV.
Since the flavor constraint $b \rightarrow s \gamma$ is particularly
prohibitive, one needs to tune the $A_0$ to obtain a WMAP satisfied 
region compatible
 with the constraints from flavor physics. In principle, one could scan over
the 4d parameter space namely $\mhalf, \mhu, \mhd, A_0$ to search for the
complete $m_{\chi_1^0}$ range in the {\it light Higgs boson} region. However,
rather than attempting a full, but very CPU - consuming scan 
over this entire parameter space, we choose to restrict ourselves to the 
subregions of the MSSM parameter space where the lightest neutralino can
be light or relatively heavier. 

Passing to Fig.\ref{a01000},
we now explore part of the MSSM parameter space where 
pair annihilation of relatively heavier neutralinos in the 
{\it light Higgs boson} zone 
can conform the WMAP data via $A$ exchange. 
%(significant contributions from $s$-channel $Z$ and 
%$t$-channel 
%neutralino exchange are also possible). 
We will see that 
the $\gamma$/$\bar P$ signals for this subset of
the MSSM parameter space, in addition to the previous results, 
can be very illustrative  to draw some general conclusions for 
the $A$ annihilation region, particularly in the context of the {\it light Higgs boson}
zone. This could be attributed to the fact that the $A$-funnel region is characterized
by large pair annihilation cross-sections (insensitive to the velocity of LSP). 
%As can be 
%
%Additionally, our interest would be 
%to see the whether this region can still produce viable $\gamma$/$\bar P$ 
%signal. 
Now, for simplicity we fix $m_0$ at the very similar value 
$m_0=600~$GeV, while
$A_0=-1000~$GeV is chosen to make $b \rightarrow s \gamma$ less restrictive.  
Here, we set  
$\mhu$ ($=2.4m_0^2$) and varied $\mhd$ ($-0.3 m_0^2<\mhd< 0.1$) 
with $\mhalf$ to obtain the WMAP-compatible regions
for neutralinos.
Our scan renders very small $\mu$ values ($150<\mu<300$), 
consequently the LSP can have large Higgsino
components. With the choice of the input parameters that we assumed
in this case, we find that the $m_{1/2}<250~$GeV region is not quite compatible
with the chargino mass limit.

%within 
%within $\mhu = 2.4m_0^2$ and $-0.3 m_0^2<\mhd< 0$.

 We can see from Fig.\ref{a01000} that 
$\mhalf \sim 300~$GeV corresponds to the {\it non-decoupling} zone for the Higgs boson.
Clearly, the lightest neutralino ($m_{\chi_1^0} \sim 120~$GeV) 
is quite far from the resonance annihilation
condition i.e., $2m_{\chi_1^0} \simeq m_A$, thus the
pair annihilation process actually occurs via off-shell $A$ 
boson exchange
 (large Higgsino component enhances the neutralino 
coupling to the Higgs boson). 
Similarly, $Z$ bosons in the $s$-channel and particularly, neutralinos
in the $t$ channel could also contribute significantly 
in the annihilation precesses. 
Thus, compared 
to the previous case, where the $b \bar b$ final state has the maximum branching
ratio, here several other final states involving the Higgs bosons ($Zh, 
ZH, W^\pm H^\pm, hA, HA, hh$) open up. We should note that 
$m_{\chi_1^0}$ cannot be much larger as then the neutralino would be 
further away from the resonance condition, which in turn makes the pair
annihilation via $A$-boson exchange less significant. Similarly, one requires 
large Higgsino components or small $\mu$ parameter to satisfy the WMAP data, 
which is not favoured in the light of recent results 
of the spin-independent neutralino-nucleon cross-section.  
Alike the previous
case, here the WMAP satisfying zone is compatible with the $b \rightarrow s \gamma$ constraint.

\subsection{Exclusion and Detectability}\label{detectability}
Before presenting our results for indirect detection, we should
define the measure, in particular what we mean by ``exclusion''
and ``detectability''.
In order to assess whether a parameter space point is excluded by current data
from Fermi or PAMELA, we should have some estimate of the background spectra 
 for both gamma-rays and antiprotons. Till the present day, and despite the remarkable
efforts devoted to the subject, no such generally accepted estimate exists.
However, and quite generically, we expect the background to be of a power law 
form $\Phi_{bkg} = a E^{b}$ in both channels. 

Now, every flux measurement contains both background as well as
(hopefully) signal events. For every parameter space point, we can compute the
gamma-ray or antiproton fluxes as described previously. Then, each point shall
be considered as excluded if there is no $(a,b)$ combination 
(i.e. the generic power-law background) for which 
the sum of the signal and the background can provide a good fit to the data.
Hence, we vary $(a,b)$, compute the corresponding backgrounds and then subsequently 
add the signal contribution
to check if there exists some background form for which
this sum provides a sufficiently good fit to the data. If no such $(a,b)$ can be found, 
the corresponding parameter space point can be considered as excluded. In practice, the criterion 
we demand is that the sum of the signal and the background should fall within the $95\%$ CL error
bars as given by the Fermi or PAMELA collaborations.

The method we follow in order to characterize a parameter space point as being 
detectable has the same philosophy:
First, we need some estimate of what future data could look like. To satisfy the criterion, 
we adopt a power
law background that is compatible with current measurements. In order to minimize the signal's significance,
we choose this background at the upper $68\%$ CL error bar of the Fermi or PAMELA experimental points.
Then, for each parameter space point, we add the signal to this background, 
creating a set of pseudo-data that could appear in the future.
As pointed out, the exact form of the background is in general unknown, but its
general form is expected to be a power-law, we could look for deviations of the pseudo-data
from such a behaviour.
If it is impossible to find a power law form that fits this pseudo-data well enough, 
then the corresponding parameter space point is characterized as detectable. If such an $(a,b)$ combination can
be found, then the signal shall be indistinguishable from the background (unless some other measurements
allow to constrain the viable $(a,b)$ combination, a possibility that we do not consider here).
The goodness-of-fit criterion, what we choose is based on the $\chi^2$ quantity, which is defined as
\begin{equation}
 \chi^2 = \sum_{i = 1}^{\mbox{\begin{tiny}nbins\end{tiny}}} 
\frac{(N_{\mbox{\begin{tiny}bkg\end{tiny}}} - N_{\mbox{\begin{tiny}exp\end{tiny}}})^2}
{N_{\mbox{\begin{tiny}bkg\end{tiny}}}},
\end{equation}
where nbins is the number of bins, taken to be $20$ in both cases, 
$N_{\mbox{\begin{tiny}exp\end{tiny}}}$ is the pseudo-data, whereas $N_{\mbox{\begin{tiny}bkg\end{tiny}}}$ is 
the background-only number of events that we try to fit to the pseudo-data.

If the best fitting power-law has a $\chi^2$ larger than $28.87$ (our problem
has $20 - 2 = 18$ degrees of freedom, since we are trying to fit $2$ variables $(a,b)$), 
this means that there is no (background-only) power-law that can fit the pseudo-data. Hence, if
$\chi^2 > 28.87$ the corresponding parameter space point is detectable, since 
the signal it generates is distinguishable from the background.

If we wish to sum up our method in ``hypothesis testing'' terms, we could say that in the case
of exclusion we are testing a null hypothesis according to which existing data can be well-described 
by dark matter annihilations plus some background form. In the detectability case, the null hypothesis
is that the pseudo-data (containing both signal and background) can be well fitted by a 
background only function.

\label{results}

\subsection{Indirect Detection results}
Our analysis shows that the current Fermi and PAMELA data do not impose any constraints
on the parameter space, at least as far as the WMAP-compliant regions are concerned.
From now on, we shall therefore stick to predictions 
concerning the detection perspectives as described in the previous paragraph.

In Fig.\ref{a01100}, with the choice of parameters described earlier, 
we also 
plot the contours where the $\chi^2$ between the background-only fit and 
the pseudo-data
becomes equal to $28.87$, a case in which the background-only hypothesis can be rejected at
$95\%$ CL and the corresponding parameter space points are thus detectable for the case $A_0 = -1100~$GeV. 
The WMAP compatible region within the
contours $D^\gamma$ or $D^{\bar P}_{Max}$ has $\chi^2 \geq 28.87$ and is, thus, 
detectable.
We assume  the Einasto profile and the MAX propagation model for gamma-rays and antiprotons respectively. The contours for
gamma-rays are blue-dashed whereas for antiprotons violet-dotted-dashed. All parameter space points lying among the 
two lines which are quasi-symmetric around the resonnance are detectable according to our previous definitions.

We note that in the case of gamma-rays, switching to another profile does not significantly alter the results.
In the case of antiprotons however, the results \textit{do} change. For the sake of clarity we omit plotting
these results, but we have calculated that both for the MIN and the MED propagation model defined previously and we find that some parts of the  
parameter space evade detection.
We note that the points that satisfy WMAP through resonnant $s$ - channel exchange of $h$
persistently evade detection, a point on 
which we shall further comment in the following paragraphs.

Before explaining these results, we notice in \cite{Baltz:2008wd} that, 
the Fermi
satelite should in principle be able to exclude WIMPs with thermal cross-sections lying in our neutralino
mass range and for $b \bar{b}$ final states. Furthermore, these estimates were extracted having in mind
the Galactic Center as observation region. As mentioned, 
looking \textit{around}
this region can further improve the observation perspectives.

Concerning our results, it is interesting that practically all the points offer quite 
good detection perspectives. This is mainly related to 
the fact that the present values of the annihilation cross-sections 
for the parameter space points satisfying the WMAP constraints
are quite high, namely of the same order as during the time of decoupling (i.e., in the thermal region). This is
mainly due to the mechanism through which the correct relic density is actually obtained. 

We already pointed out that in
this scenario the mechanism that drives neutralino annihilation is resonnant 
$s$-channel pseudoscalar Higgs boson
exhange, apart from the small region at low $m_{1/2}$ and relatively large 
$m_A$, where the dominant mechanism is CP-even light
Higgs exchange. In the case of annihilation through an $A$ propagator, the cross-section is practically insensitive to
velocity changes as pointed out for example in \cite{dmreview}. This leads to the conclusion that the
self-annihilation cross-section stays quite high even at present times. It is really instructive to compare
this regime with the corresponding points where the acceptable 
relic density is produced via neutralino pair annihilation into
$h$ boson. In this case, $<\sigma v>$ tends to zero as the LSP velocity does so 
(i.e., at present times, which is relevant
for indirect detection experiments). 

This is indeed an interesting
effect, which renders the $h$-pole points practically invisible to indirect detection experiments. The same
has been pointed out, for example, in \cite{Bernal:2009jc}, where detectability limits (although defined
differently) seem to systematically ``avoid'' the $h$-pole region. Now, we 
should note that
the decay modes in the funnel region are dominated by the $b \bar{b}$ final 
states. 
This is due 
to the fact that neutralino pair annihilation into $A$ and then into down 
type fermions 
is proportional to $\tanb$ along with the quark masses. It is well-known 
that the ($b \bar b$) final state
yielding relatively rich photon spectra if compared, for example, to the leptonic case. 
 On the other hand for the antiproton yield, the decays of 
light quarks have the tendency of
producing more antiprotons than $b \bar{b}$ pairs.

Let us proceed to our second scenario, i.e., Fig.\ref{a01000}. Once again, 
the blue-dashed line depicts region where $\chi^2 = 28.87$ for gamma-rays whereas the 
violet-dotted-dashed line represents
the same condition for antiprotons. Astrophysical assumptions are the same as in the previous case. Points lying 
above, below or on the left of the contours are detectable. If we plotted the gamma-ray results for the
other two profiles we examined, results would be practically unchanged. In the case of the two other
propagation models for antiprotons, AMS-02 will be blind to the part of the relic density satisfying points.
We can see that in this scenario, the 
perspectives are also quite good. We should note that we are still lying 
in the $A$-pole region (with significant contribution from $s-$channel $Z$ and $t-$channel 
neutralino exchange): The neutralino annihilation
is driven by the $s$-channel pseudoscalar exchange. Once again, $<\sigma v>$ lies roughly in the 
typical thermal region. But in this case, the lightest neutralino has a higher mass than previously.
This is the reason why for relatively large values of $m_{1/2}$, we have a certain
deterioration in the detection perspectives, particularly in the lower branch of the WMAP
compliant parameter space. 
This is mostly visible in the antiproton channel, 
where we see that practically all LHS points are invisible at AMS-02. We have checked that if we
consider more stringent gamma-ray detectability criterion, the same tendency would be visible for the
corresponding contour as well. This behaviour could be connected to the fact that
in this particular parameter space (lower branch), 
Higgs and gauge bosons have significant branching fraction in the final state. 
This is not the case
for the upper branch, where the dominant channel is $b \bar b$.
The main effect
of a final state including Higgses is to shift the energy spectrum towards lower energies
(since we consider the Higgs bosons to decay predominantly into $b\bar{b}$ pairs), where the
background is larger. It is thus more difficult to disentagle the non-power law component of the
spectrum (i.e. the signal) from the background.

In the two scenarios we examine two regimes, one with a quite light neutralino 
($50 \lesssim m_{\chi_1^0} \lesssim 80$ GeV) and one with a relatively heavier one
($100 \lesssim m_{\chi_1^0} \lesssim 130$ GeV), finding that a good part of our parameter
space should be visible at Fermi and AMS-02. There is, 
however, one question that could be asked, namely what happens in the intermediate mass regime, 
particularly in the context of {\it light Higgs boson} zone.  
The answer could be given once more by considering that it is well-known that
increasing the WIMP mass tends to aggravate detection perspectives, if the same final states and
self-annihilation cross-section are assumed. Both the $Br_i$'s and $\left\langle \sigma v \right\rangle$
remain quite stable in value from lighter to higher masses in our model: The final state is mostly $b \bar{b}$ 
(in the second case Higgs final states are also significant which subsequently decays mostly 
into $b \bar b$)
and the cross-section is of the typical thermal value, both during decoupling \textit{and} at present times.
Thus, it is easy to infer that the intermediate mass regime would also be able
to produce rich $\gamma-$ray or anti-proton signals.
Overall, this $A$-pole scenario that we have examined
can be considered as quite promising for indirect detection. We shall further comment on that 
in the Conclusions section.

%%%%%%%%%%%%%%%%%%%%%%%%%%%%%%%%%%%%%%%%%%%%%%%%%%%%%%%%%%%%%%%%%%%%%%%%%%%%%%%%
%%%%%%%%%%%%%%%%%%%%%%%%%%%%%%%%%%%%%%%%%%%%%%%%%%%%%%%%%%%%%%%%%%%%%%%%%%%%%%%%%%%%%%%%%%%%%%%%%%%%%%%
%%%%%%%%%%%%%%%%%%%%%%%%%%%%%%%%%%%%%%%%%%%%%%%%%%%%%%%%%%%%%%%%%%%%%%%%%%%%%%%%%%%%%%%%%%%%%%%%%%%%%%%
%%%%%%%%%%%%%%%%%%%%%%%%%%%%%%%%%%%%%%%%%%%%%%%%%%%%%%%%%%%%%%%%%%%%%%%%%%%%%%%%%%%%%%%%%%%%%%%%%%%%%%%
\section{Discussion on the constraints from Direct Detection experiments}
\label{sec:DirectDetection}
\begin{figure}[tb!]
\begin{center}
\includegraphics[width=0.50\textwidth,clip=true,angle=-90]{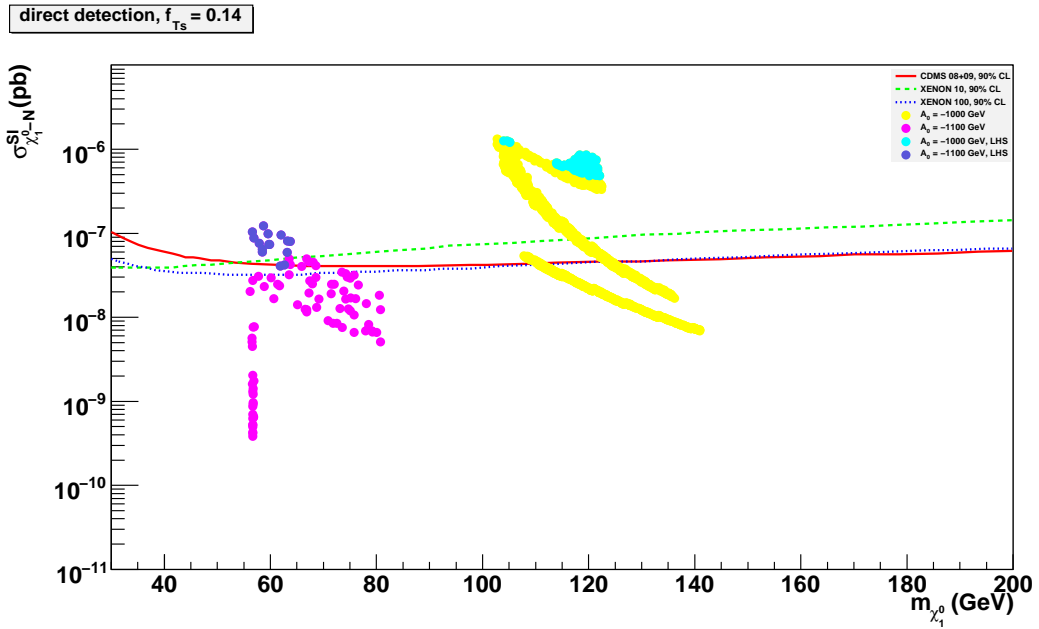}
\end{center}
\caption{
$(m_{\chi^0_1}, \sigma_{\chi^0_1-N}^{SI})$ combinations 
along with the relevant exclusion limits from direct detection
experiments. The points lying above the lines are in principle excluded according to the published
limits. Yellow points correspond to a trilinear coupling value of 
$A_0 = -1000~$GeV whereas pink ones to
$A_0 = -1100~$GeV. The light blue and  the dark blue points represent the
{\it light Higgs boson} regime for the two scenarios respectively. 
$f_{Ts}^{(p, (n))}$ is taken at the default DarkSusy value, namely $0.14$.}
\label{DirectLimits}
\end{figure}

In Fig.\ref{DirectLimits} we depict the WMAP-compliant parameter
points on the $(m_{\chi^0_1}, \sigma_{\chi^0_1-N}^{SI})$ (neutralino mass - neutralino-nucleon spin-independent
scattering cross-section) plane and compare them against the three strongest bounds available in the 
literature: The combined 2008 and 2009 CDMS-II \cite{cdms-2} results, the constraints from the XENON10 experiment as
well as the latest bounds from XENON100. We take the two former ones from ref. \cite{Kopp:2009qt} and the
latter from \cite{xenon100}. We further highlight the points falling 
into the light Higgs scenario
defined throughout this paper with different colors.

We indeed see that many of our points fall largely within the region that 
is supposed to be excluded
from the existing data. Whereas the $A_0 = -1100~$GeV scenario (pink points) 
is more or less satisfying the constraints,
the scenario with $A_0 = -1000~$GeV (yellow points) is in most cases largely above the 
limits, exceeding 
by more than an order of magnitude compared to the CDMS-II and XENON100 allowed cross-sections. The large values for $\sigma_{\chi^0_1-N}^{SI}$ in the latter case can be 
attributed 
to the large Higgsino components of the neutralino which enhance the coupling
${\cal{C}}_{\chi_1^0 \chi_1^0 h(H)}$. 
In both cases, 
the points belonging to the {\it light Higgs boson} region
fall within the excluded zones.

However, it has been repeatedly pointed out in the literature that there can be significant uncertainties
that complicate the assertion on whether a particular model is excluded or not
 \cite{DDMe}. More specifically:
\begin{itemize}
 \item Uncertainties can arise in the calculation of the neutralino-nucleon elastic scattering cross-section,
which can be due to a number of factors. For example, as described in
detail in \cite{Ellis:2008hf}, significant uncertainties can arise in the passage from the parton-level cross-section
to the hadronic level one. 
 \item Some uncertainties might be present in the passage from the hadronic to the nuclear level. Indeed, 
at the end of the day the primarily constrained quantity is the WIMP-\textit{nucleus} elastic scattering cross-section
and not the WIMP-\textit{nucleon} one. 
%{\bf In fact the nuclear form factors particularly related to the strange quark is the major source of uncertainties.} 
%It would not thus be impossible to think that the nuclear form factors could be a source of some error.
 \item The local dark matter density is by no means a perfectly well known quantity and is in fact a normalization
factor in the overall procedure of computing the WIMP-nucleus scattering rate. This uncertainty 
also enters in the indirect detection calculation, in fact in a more severe way since 
the gamma-ray or antiproton fluxes
depend quadratically on the aforementioned quantity.
%halo profile normalization for annihilating dark matter. 
%Although such an analysis
%is omitted in the present work, this point should be kept in mind in the following.
 \item Little is known on the true velocity distribution of the WIMPs in the detector rest frame as well as on the
escape velocity at which the integral over the velocity distribution should be truncated.
\end{itemize}
The first point concerns our calculation of the spin-independent
neutralino-nucleon elastic scattering cross-section. 
In other words, and refering to Fig.\ref{DirectLimits}, we 
expect that a certain variation in the position of parameter space points on the $(m_{\chi^0_1}, \sigma_{\chi^0_1-N}^{SI})$ plane
should be allowed. 
The other remarks apply to the experimental limits published
from the various collaborations, i.e., they can amount to a change in the position of the exclusion lines. 

\begin{figure}[tb!]
\begin{center}
\includegraphics[width=0.50\textwidth,clip=true,angle=-90]{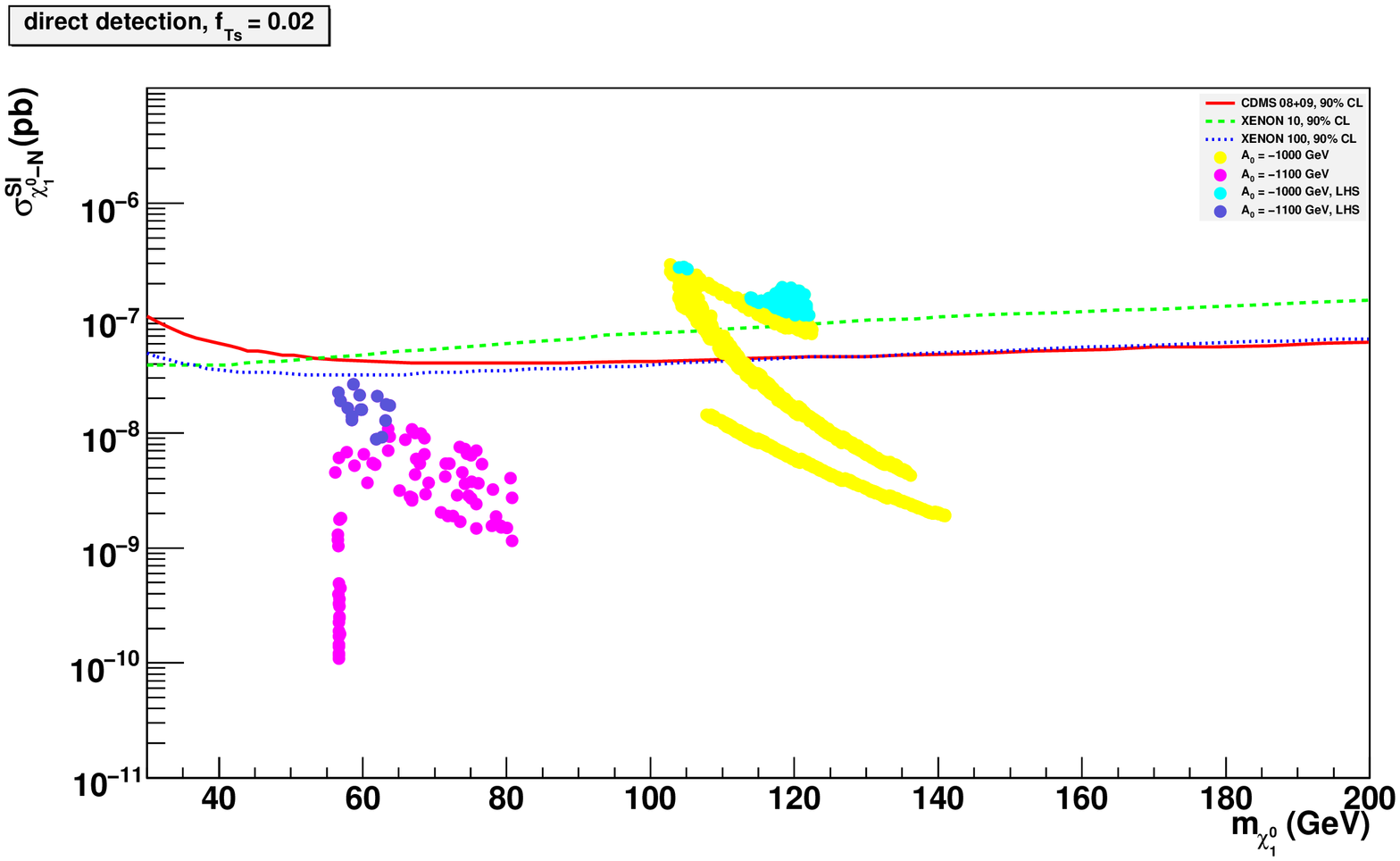}\\
\includegraphics[width=0.50\textwidth,clip=true,angle=-90]{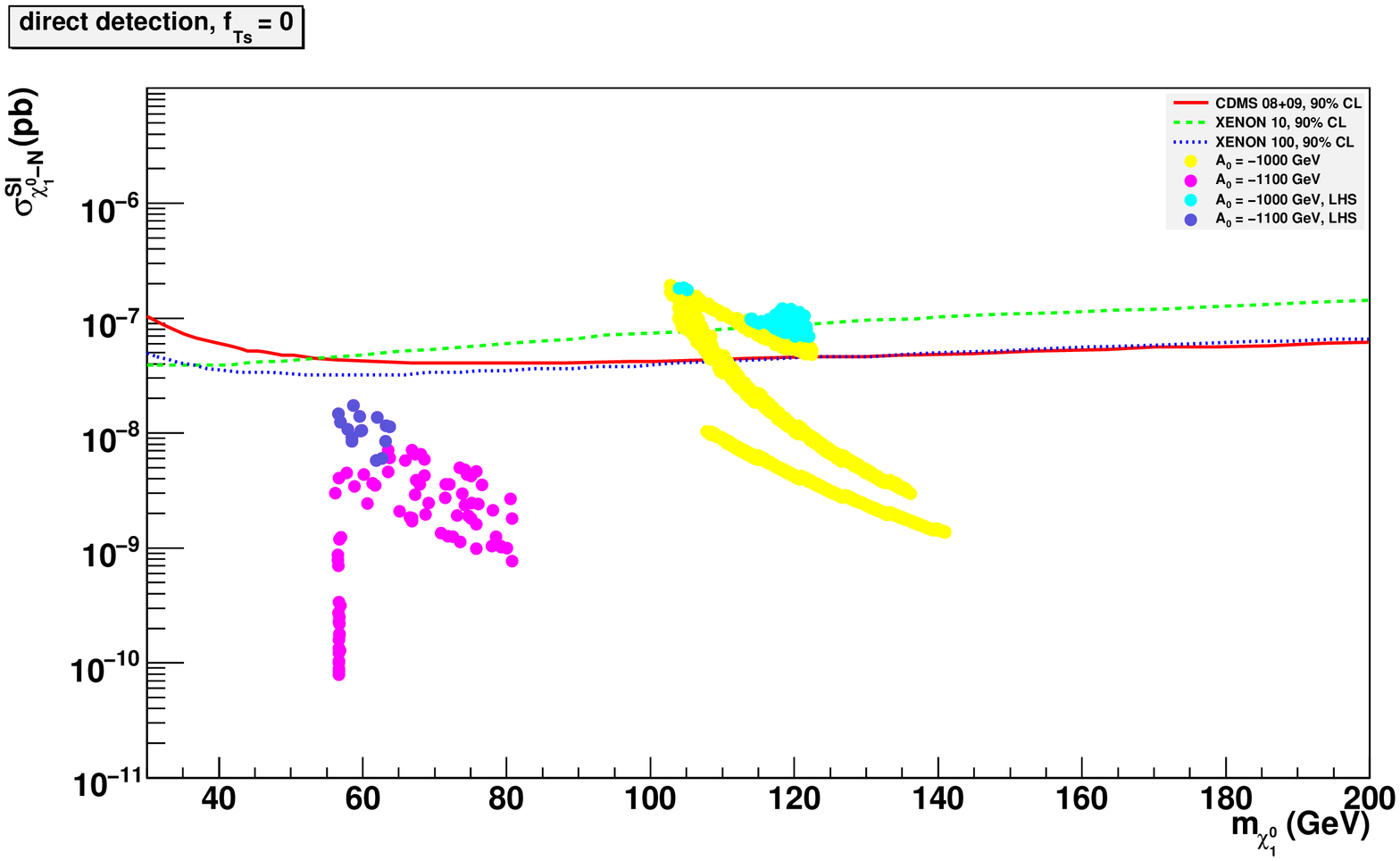}
\end{center}
\caption{
As in fig.\ref{DirectLimits} but for $f_{Ts}^{(p, (n))} = 0.02$ (top) and $0$ (bottom).}
\label{DirectLimitsfTs}
\end{figure}

In ref. \cite{Ellis:2008hf}, a systematic study of the hadronic uncertainties entering the neutralino-nucleon
scattering cross-section is performed. It turns out that the most striking and influential uncertainty
comes from the pion-nucleon $\sigma_{\pi N}$ term 
%related to
%the strange quark content of the nucleon (see Sec.\ref{dir_detection})
which is poorly known but an essential ingredient
for a precise calculation of the relevant cross-section. This source of uncertainty alone
can give rise to a variation in the spin-independent cross-section of more than an order of magnitude \cite{Cao:2010ph,Ellis:2008hf}. This means
that the relevant neutralino-nucleon scattering cross-sections that we 
have calculated can in fact vary by
a factor of more than $10$. As mentioned in Sec.\ref{dir_detection}, uncertainty in 
$\sigma_{\pi N}$ may decrease the coefficient $f_{Ts}^{(p, (n))}$ and this is in fact predicted by recent
lattice simulation.
Thus, we study the variation in $\sigma_{\chi^0_1-N}^{SI}$ 
with smaller $f_{Ts}^{(p, (n))}$ values. 
Following the discussion in Sec.\ref{dir_detection}, 
here we consider two 
representative values for $f_{Ts}^{(p, (n))}$ namely $0.02$ and $0$. We present our results in 
fig.\ref{DirectLimitsfTs}. Indeed, we can
see that the neutralino-nucleon spin-independent cross-sections decrease by significant factors, reaching up to an order of magnitude (particularly for $f_{Ts}^{(p, (n))}=0$). 
This clearly starts raising questions on whether a good portion of the parameter
space is excluded (as one would naively expect from Fig.\ref{DirectLimits}) 
or not.

We see, however, that - especially in the heavier neutralino scenario - there are still some
parameter points lying above the exclusion lines (roughly by a factor of 
$2 - 3$). This is particularly true for the LHS scenario. However, this conclusion may become weaker
if one considers the other uncertainties which we describe below.
%Are these points definitively excluded? The answer is, once again, not that straightforward.

Passing to the nuclear level requires modelling of the nucleon density within the nucleus. The most commonly
used parametrization is the one presented by Engel in \cite{Engel:1991wq}. 
Now, the 
deviation
that might arise from different form factor parametrizations has been performed, for example, in
 \cite{Duda:2006uk}, where the authors find that the exclusion lines can shift vertically by roughly a
factor of $1.2$.

Next, we discuss the uncertainty related to the local dark matter density. 
Most experimental analyses 
including the ones considered in the present work 
are performed assuming
the so-called standard halo model. In this model, the local DM density is usually taken to have a fixed value 
of $0.3$ GeV cm$^{-3}$. Typically, the uncertainties associated with the local density value 
are given to be of the order of $10\% - 20\%$ at a $68\%$ confidence level, 
i.e., less than a factor of two. 
It is, however, true that these estimates depend on specific astrophysical assumptions.
There have been authors who using a wide range of possible halo profiles  \cite{Weber:2009pt} or no
halo modelization at all \cite{Salucci:2010qr} find values for the local density ranging
from $0.2$ up to roughly $0.5$ GeV cm$^{-3}$ at $68\%$ CL.
Furthermore, clumpy behaviour could further disrupt these estimates, although $N$-body
simulations seem to rather disfavour such a possibility \cite{Vogelsberger:2008qb}. 
Thus there may be a factor $\sim 2$ variation in the local
DM density, which would rescale the corresponding WIMP-nucleon scattering cross-section limits by the
same amount.

A recent treatment of the astrophysical uncertainties involved in direct DM detection is \cite{McCabe:2010zh}, 
while a summary of some of these uncertainties with several references can also be found in \cite{Green:2010ri}.
Varying the escape velocity can play an important role either for 
larger LSP masses which are irrelevant in our case
 or for models involving quite different processes than the ones involved in the MSSM neutralino
case \cite{weiner}. The same applies to the uncertainties arising from the experimental error in the determination of the 
Sun's circular velocity around the galactic center, which assumes a 
Maxwell-Boltzmann velocity distribution 
for neutralinos. Therefore we expect only a small modifications in the exclusion limits ($\sim
O(10\%)$) due to these aforementioned variations.

We see that overall, and despite the apparent exclusion of a large portion 
of LHS points, there is still
quite some margin for changing the relation among the predicted $(m_{\chi^0_1}, \sigma_{\chi^0_1-N}^{SI})$ as derived
from the model and the exclusion limits as presented by the corresponding collaborations. We feel it is
reasonable to say that we cannot assess that easily whether the parameter space points lying above the exclusion
lines in Fig.\ref{DirectLimits} are actually excluded or not.

We should clarify at this point that the previous remarks have by no means the purpose of demeaning the
remarkable works that are done both by theorists and experimentalists in order to develop tools for calculations
and extract reliable bounds. Our goal was just to illustrate that it might still be
meaningful to examine models which at first sight appear to be excluded. This becomes particularly apparent
from our calculation of the spin-independent cross-section for different values of $f_{Ts}^{(p, (n))}$.

%%%%%%%%%%%%%%%%%%%%%%%%%%%%%%%%%%%%%%%%%%%%%%%%%%%%%%%%%%%%%%%%%%%%%%%%%%%%%%%%%%%%%%%%%%%%%%%%%%%%%%%
%%%%%%%%%%%%%%%%%%%%%%%%%%%%%%%%%%%%%%%%%%%%%%%%%%%%%%%%%%%%%%%%%%%%%%%%%%%%%%%%%%%%%%%%%%%%%%%%%%%%%%%
%%%%%%%%%%%%%%%%%%%%%%%%%%%%%%%%%%%%%%%%%%%%%%%%%%%%%%%%%%%%%%%%%%%%%%%%%%%%%%%%%%%%%%%%%%%%%%%%%%%%%%%
\section{Conclusions}
\label{sec:con}
We have examined the dark matter - related phenomenology of a 
particular subset of the MSSM parameter space adapting 
nonuniversal Higgs masses at the GUT scale. 
We confine ourselves to a particular regime where the MSSM Higgs scalars 
as well as neutralinos can be relatively light.
Our principal investigation is devoted where lightest CP-even Higgs boson is 
below the LEP2 limit, but still allowed  
due to the reduced value of $Z$$Z$$h$ coupling. 

The WMAP limits on the dark matter relic density are satisfied 
by the neutralino 
self-annihilation through $A$, $h$ or $H$ propagators. For illustration
purposes we have chosen two specific examples with different neutralino masses.
Then we have examined the detectability of the scenarios
in the current Fermi-LAT and the upcoming AMS-02 experiment for the detection of gamma-rays and
antiprotons respectively.

We have found that overall the model offers quite promising
detection prospects: The self-annihilation
cross-section at zero velocity, relevant for indirect detection,
 is quite high, i.e., of the same order of
magnitude as needed
in order to get the correct relic density for the WIMPs.
This is especially the case when annihilation is driven by $s$-channel pseodoscalar exchange. 
Furthermore, the final 
states being mostly comprised of $b \bar{b}$ pairs (with significant Higgs
bosons final states for relatively larger neutralino masses), the 
resulting spectra are quite rich both
in gamma-rays and antiprotons (especially for photons). 
In this context we did not
consider the possibility of substructure in the galactic halo, which could enhance the signal 
by roughly a factor of $10$ as is well-known in the literature.

Finally, we have computed the neutralino-nucleon spin-independent 
scattering cross-section and find that under the most common assumptions
a significant part of the LHS region seems to be excluded by current direct detection
measurements. We however showed that numerous uncertainties could severely influence this assessment.

Furthermore, the preceeding analysis demonstrates, once more, the value of a 
multi-messenger approach towards dark matter detection. Firstly, each detection mode offers
a distinct probe for dark matter signals which can be used to cross-check others. Then, 
some regions of the parameter space might be undetectable in some channel, but visible in some
other: This is the case of the LHS in our second benchmark, which is invisible in antiproton
detection but visible in gamma-rays or direct detection. Overall, different detection techniques can act in a highly
complementary way.

We would like to close this article with a remark that seems to have a 
more general validity than
the model we examined in our case. As we said, the 
regions where the correct relic density can
be obtained in the MSSM are quite limited in their general characteristics. 
One of them is
the so-called funnel region, where the 
self-annihilation cross-section is enhanced through 
resonnant $s$ - channel exchange of Higgses, $h$, $H$ and $A$. 
In the latter case, it is known for quite some time that 
the cross-section does not depend strongly on the neutralino velocity. 
The fact that the Large Area Telescopes being at least as powerful as 
Fermi can probe self-annihilation cross-section values of the order of 
$3 \cdot 10^{-26}$ cm$^3$ sec and masses of roughly up to $200$ GeV has 
also been demonstrated. It can hence be seen that the $A$ - 
pole region for reasonable neutralino
masses constitutes one of the most favoured regions for indirect detection
in both the gamma-ray and the anti-proton channel.
 On the contrary, other WMAP compatible regions namely
the $h$-pole resonance annihilation region or the coannihilation regions 
(where the Higgsino component is tiny)
 where the cross-section is much smaller at present
times than during decoupling cannot produce so large indirect detection
rates. 

%%%%%%%%%%
\section*{Acknowledgements}
This work has been done partly under the ANR project TAPDMS No {\bf 
ANR-09-JCJC-0146}.
D.D. acknowledges
support from the Groupement d'Int\'er\^et Scientifique P2I.
The work of AG and YM are supported in part by the E.C. Research Training Networks under contract {\bf MRTN-CT-2006-035505}.
The authors would like to thank the anonymous referee for valuable comments and remarks.

%%%%%%%%%%%%%%%%%%%%%%%%%%%%%
%\bibliographystyle{utphys}
%\bibliography{biblio}

\end{document}